\documentclass[amsmath,amssymb,twocolumn,showpacs]{revtex4}

\usepackage{mathptmx}

\usepackage{graphicx}
\usepackage{bm}
\usepackage{amsthm}
\usepackage{eepic}

\newcommand{\R}{\mathbb{R}}
\newcommand{\Z}{\mathbb{Z}}

\newcommand{\tr}{\mathop{\textup{tr}}}
\newcommand{\vol}{\mathop{\textup{vol}}}

\theoremstyle{plain}
\newtheorem{theorem}{Theorem}

\newtheorem{lemma}[theorem]{Lemma}
\newtheorem{proposition}[theorem]{Proposition}

\begin{document}

\title{Ground states and formal duality relations in the
Gaussian core model}

\author{Henry Cohn}
\affiliation{Microsoft Research New England, One Memorial Drive,
Cambridge, Massachusetts 02142, USA} \email{cohn@microsoft.com}

\author{Abhinav Kumar}
\affiliation{Department of Mathematics, Massachusetts Institute of
Technology, Cambridge, Massachusetts 02139, USA}
\email{abhinav@math.mit.edu}

\author{Achill Sch\"urmann}
\affiliation{Institute of Applied Mathematics, TU Delft, Mekelweg 4,
2628 CD Delft, The Netherlands} \email{a.schurmann@tudelft.nl}

\date{December 14, 2009}

\begin{abstract}
We study dimensional trends in ground states for soft-matter systems.
Specifically, using a high-dimensional version of Parrinello-Rahman
dynamics, we investigate the behavior of the Gaussian core model in up
to eight dimensions. The results include unexpected geometric
structures, with surprising anisotropy as well as formal duality
relations. These duality relations suggest that the Gaussian core model
possesses unexplored symmetries, and they have implications for a broad
range of soft-core potentials.
\end{abstract}


\maketitle

\section{Introduction}

Soft-matter systems are notoriously difficult to analyze theoretically,
and much of what we know about their phase diagrams is based on
numerical simulations.  Even classical ground states can almost never
be derived from first principles (see Refs.\ \cite{T,Su1,Su2,L,CK1} for
some rare exceptions).  In this article, we place phenomena such as
crystallization and solid-solid phase transitions in a broader context
by studying dimensional trends in the Gaussian core model \cite{S}, in
which particles interact via a Gaussian pair potential.

The Gaussian potential models the entropic effective interaction
between the centers of mass of polymers \cite{FK}, and it is one of the
simplest and most elegant soft-core potentials.  The behavior of the
Gaussian core model in two and three dimensions is relatively well
understood (see, for example, Refs.~\cite{LLWL} and \cite{PSG}), but in
higher dimensions, it remains mysterious.  Dimensions above three are
an excellent test case for the study of phenomena such as decorrelation
\cite{TS1}, and this fits into the long tradition in statistical
mechanics of studying the effect of dimensionality in interacting
systems, such as critical dimensions for mean-field behavior (see
Section~16.7 in Ref.~\cite{H} for an overview).

Furthermore, higher dimensions play a fundamental role in information
theory. Sphere packing is the low-density limiting case of the Gaussian
core model, and sphere packings are error-correcting codes for a
continuous communication channel.  The dimension of the ambient space
for the packing depends on the channel and coding method used, and it
can be quite high in practice \cite{Mac} (up to thousands of
dimensions). Thus, coding theory is a powerful motivation for the study
of the high-dimensional Gaussian core model.

Our conclusions are based on molecular dynamics simulations \cite{FS}.
Such simulations are frequently used and often highly informative, but
the computational difficulties are immense for many-body systems.
Thanks to the curse of dimensionality \cite{B}, high-dimensional
simulations typically require exponentially many particles, which
severely limits the range of dimensions in which simulations are
possible. To address this problem, we use a high-dimensional version of
Parrinello-Rahman dynamics \cite{PR1,PR2}. Instead of imposing periodic
boundary conditions using a fixed background lattice, we dynamically
update the lattice using the intrinsic geometry of the space of
lattices.  By increasing the adaptivity of the simulation, we are able
to minimize the number of particles and avoid unnecessary computational
complexity. This lets us carry out higher-dimensional simulations than
were previously possible.

In this article we carry out Parrinello-Rahman simulations of the
Gaussian core model in dimensions two through eight.  In addition to
observing surprising geometrical phenomena such as anisotropy, we find
formal duality relations between Gaussian core ground states at
densities $\rho$ and $1/\rho$. Although such duality is known between
reciprocal Bravais lattices (see, for example, Ref.~\cite{TS2}), it
rarely holds for other structures, and its occurrence here suggests a
deeper, not yet understood symmetry of the Gaussian core model itself.

Our approach fits into a program pioneered by Gottwald \emph{et al.\/}
\cite{GKL} and applied in Refs.~\cite{GLKL1,GLKL2}. They use genetic
algorithms to search the space of candidate structures into which a
fluid can freeze. Our goals are similar, but we make use of more
analytic tools. Specifically, we compute gradients in the space of
lattices, which enables us to use more powerful optimization techniques
such as gradient descent or conjugate gradient.

\section{Framework}

Consider a periodic configuration of particles in $n$-dimensional
Euclidean space $\R^n$. Such a configuration is specified by an
underlying Bravais lattice $\Lambda \subset \R^n$, together with a
collection of translation vectors $v_1,\dots,v_N$.  In crystallographic
terms, it is a lattice with basis.  The particles are located at the
points $x + v_i$ for $x \in \Lambda$ and $1 \le i \le N$.

Given a radial pair potential $V$, the average energy per particle is
$$
\frac{1}{2N} \sum_{i=1}^N \sum_{j=1}^N
\sum_{\shortstack[c]{$\scriptstyle z \in \Lambda$\\
\scriptsize $z \ne 0$ if $i = j$}} V(|z+v_i-v_j|).
$$
This quantity is the potential energy of the system, and the
configuration is a classical ground state if it minimizes potential
energy, even allowing $\Lambda$ and $N$ to vary but keeping the
particle density fixed.  [The density equals $N/\det(\Lambda)$, where
$\det(\Lambda)$ is the absolute value of the determinant of a basis for
$\Lambda.$] In other words, classical ground states correspond to the
canonical ensemble at zero temperature.

Simulations typically fix $\Lambda$ and allow $v_1,\dots,v_N$ to vary.
This amounts to using $\Lambda$ to define periodic boundary conditions.
The lattice $\Lambda$ remains the same throughout this process, and it
is often chosen to be proportional to a hypercubic lattice $\Z^n$ for
computational simplicity.  This imposes artificial structure on the
system, and $N$ must be chosen quite large to minimize the effects of
this structure. For example, if one wants the lattice spacing in
$\Lambda$ to be an order of magnitude larger than the typical spacing
between particles, then $N$ must grow roughly like $10^n$.  For large
$n$, this is clearly infeasible, and even for $n=6$, it requires
careful use of all available computational improvements.  Many
simulations, such as those in Ref.~\cite{SDST}, are therefore limited
to roughly six dimensions.

By allowing $\Lambda$ to vary, one might hope to use a much smaller
value of $N$.  In the most extreme case, one could take $N=1$ and study
all Bravais lattice configurations.  The naive dynamics then fail
completely, because the forces in a Bravais lattice balance perfectly.
Nevertheless, the potential energy varies dramatically between
different Bravais lattices, with corresponding dynamics on the space of
lattices. The Parrinello-Rahman method uses these dynamics.  It can
therefore simultaneously update the underlying Bravais lattice
$\Lambda$ and the particle locations $v_1,\dots,v_N$.

Before describing the simulation results, we will give a derivation of
the $n$-dimensional version of Parrinello-Rahman dynamics.  It is
equivalent to the original formulation in Refs.~\cite{PR1,PR2}, except
of course for the change in dimension.  Instead of deriving it from a
postulated Lagrangian, we show how it follows naturally from the
intrinsic geometry of the space of lattices.  Presenting the derivation
gives us an opportunity to describe some of the computational issues
that become important in higher dimensions, such as the use of lattice
basis reduction algorithms.

\section{Geometry of the space of lattices}

We will represent Bravais lattices by positive-definite, symmetric
matrices. Specifically, there is a linear transformation with matrix
$T$ such that $\Lambda = T \Z^n$.  If $v=Tw$, then the squared vector
length $v^t v$ (the exponent $t$ denotes transpose and we use column
vectors) is $w^t T^t T w$.  Set $G = T^t T$.  This Gram matrix
represents the metric in coordinates in which the underlying lattice is
$\Z^n$. Instead of fixing the metric and deforming the lattice, we will
fix the lattice and deform the metric.  This approach is used to define
the intrinsic geometry on the space of lattices (see, for example,
Ref.~\cite{Sch}), and it makes the formulas quite a bit simpler.

To simplify the notation, define the function $f$ of squared distance
by $f(s) = V(\sqrt{s})/2$. Furthermore, write the vectors
$v_1,\dots,v_N$ in the new coordinates as $v_i = T u_i$. Now the
potential energy of the system  is
$$
U(G) = \frac{1}{N}\sum_{w} f(w^t G w),
$$
where we sum over all vectors $w$ of the form $u_i-u_j+x$ with $1 \le
i,j \le N$,  $x \in \Z^n$, and $x \ne 0$ if $i=j$.

The gradient of this sum as a function of $G$ equals the matrix
$$
\nabla U(G) = \frac{1}{N}\sum_{w} f'(w^t G w) w w^t.
$$
To see why, note that if we vary the $i,j$ component $G_{ij}$ of $G$
while leaving all other entries fixed [and write $w=(w_1,\dots,w_n)$],
we find that
$$
\frac{\partial}{\partial G_{ij}} f(w^t G w) = f'(w^t G
w) w_i w_j.
$$

When we update the configuration, we must fix $\det(G)$, so that the
density of the system does not change [note that $\det(G) =
\det(\Lambda)^2$]. However, the gradient does not respect this
constraint.  Instead, we must use the modified gradient
$\widetilde{\nabla}U(G)$ conditioned on fixing the determinant, which
is computed as follows. If we define the standard inner product
$\langle \cdot,\cdot \rangle$ on the space of symmetric matrices by
$\langle A,B \rangle = \tr(AB)$, then to preserve $\det(G)$ we must
remove the component of $\nabla U (G)$ in the direction of $G^{-1}$,
because
$$
\det(G +
\varepsilon \nabla U(G)) = \det(G) (1 + \varepsilon  \langle G^{-1},
\nabla U(G)\rangle + O(\varepsilon^2)).
$$
We define the modified gradient $\widetilde{\nabla} U(G)$ by
$$
\widetilde{\nabla} U(G) = \nabla U (G) - \frac{\langle
\nabla U(G), G^{-1} \rangle} {\langle G^{-1}, G^{-1} \rangle} G^{-1},
$$
so that $\langle G^{-1}, \widetilde{\nabla}U(G) \rangle = 0$.

We could avoid this last complication by replacing the canonical
ensemble with the grand canonical ensemble and controlling the particle
density via the chemical potential.  However, the modified gradient is
not difficult to use, and our approach is convenient if one wishes to
target a specific density.

\section{Parrinello-Rahman dynamics}

The lattice gradient has two components, one for changing $G$ (computed
above as the modified gradient) and one for changing $u_1,\dots,u_N$.
For the latter, if we write $w = u_i-u_j+x$, then we need the gradient
of $f(w^t G w)$ as a function of $u_i$ and $u_j$, which is easily
computed as follows. With $w = u_i-u_j+x$, the gradient of $f(w^t G w)$
as a function of $u_i$ is $2 f'(w^t G w) G w$. Thus, the $u_i$
component of the gradient of potential energy is the sum of $(2/N)
f'(w^t G w) G w$ over all $w$ of the form $u_i-u_j+x$ for some $j$ plus
the sum of $-(2/N) f'(w^t G w) G w$ over all $w$ of the form $u_j - u_i
+ x$ for some $j$. The full lattice gradient is made up of both the $G$
and the $u_1,\dots,u_N$ components.

Parrinello-Rahman dynamics consists of using the lattice gradient to
define forces on the configuration. Thus, $G$ changes as well as
$u_1,\dots,u_N$. The simplest application is gradient descent, where we
seek a local minimum for energy by following the negative gradient (or,
better yet, using the conjugate gradient algorithm). Of course, one
could also use the forces in the usual way to define accelerations
rather than simply velocities, but we will focus on gradient descent
here because of our interest in ground states.

The infinite sums in the algorithm must be truncated in practice, by
summing only over the $w$ such that $w^t G w$ is at most some bound
(chosen based on the decay rate of $f$). Writing $w = u_i-u_j+x$, we
must enumerate all $x \in \Z^n$ with this property. To do so, we use
the Fincke-Pohst algorithm \cite{FP}, which is far more efficient than
brute-force searches.  By contrast, simply summing over all the vectors
in a large box becomes exponentially inefficient in high dimensions. We
also periodically apply the $\textup{L}^3$ lattice basis reduction
algorithm \cite{LLL}, which changes basis so as to keep the entries of
$G$ small, and we renormalize $G$ to maintain a constant determinant
(so that small numerical errors do not accumulate).

As the simulation progresses, the metric changes and the connection
with the original coordinates is lost.  Nevertheless, using the
Cholesky decomposition \cite{D}, we can recover the Bravais lattice $T
\Z^n$ with basis $Tu_1,\dots,Tu_N$ from the matrix $G$ and the vectors
$u_1,\dots,u_N$ by finding $T$ such that $G = T^t T$.

Because of the need to use tools from lattice basis reduction theory
and linear algebra, the Parrinello-Rahman method is not as simple to
implement in high dimensions as more straightforward methods are.
However, in compensation it adapts itself to the structure of the
system being considered and can therefore provide improved results.
Furthermore, it is compatible with other standard computational methods
such as Ewald summation or fast multipole methods \cite{FS}.

\section{Results for the Gaussian core model} \label{section:results}

For the Gaussian core model, we take $V(r) = \exp(-\pi r^2)$ and hence
$f(s) = \exp(-\pi s)/2$.  The choice of the constant $\pi$ amounts to
fixing the length scale, and it is chosen to make $V$ self-dual under
the Fourier transform.  Let $\rho$ denote the particle density.

The ground states of this model have been thoroughly examined in up to
three dimensions, although except in $\R^1$, no proof is known
\cite{CK1}. In $\R^2$, at all densities the ground state is the
triangular lattice $A_2$. In $\R^3$, the face-centered cubic lattice
$D_3$ is optimal at low densities, and the reciprocal body-centered
cubic lattice $D_3^*$ is optimal at high densities \cite{SW}. The
cross-over point is at $\rho = 1$, but in fact the Maxwell
double-tangent construction (i.e., the convexity of potential energy as
a function of $1/\rho$) leads to phase coexistence for
$0.99899854\ldots \le \rho \le 1.00100312\ldots$. This appears to give
a complete description of the phase transition from $D_3$ to $D_3^*$.

Little is known in higher dimensions, despite the connections with
coding and information theory. Cohn and Kumar \cite{CK1} conjectured
that the $E_8$ and Leech lattices are universally optimal when $n=8$ or
$24$, respectively.  (In other words, they are ground states for the
Gaussian core model at all densities. As shown in Ref.~\cite{CK1}, this
implies optimality for many other potentials, such as all inverse power
laws.) Torquato and Stillinger \cite{TS2} conjectured that in at most
eight dimensions, certain Bravais lattices are always optimal at
sufficiently high or low densities, but their conjecture was disproved
in five and seven dimensions \cite{CK2}. Despite extensive exploration
\cite{ZST}, the true ground states have remained a mystery.  Because of
the difficulty of simulation, previous studies have made use only of
structures already known for other reasons. Comparing such structures
in the Gaussian core model is of course of value, and it can sometimes
lead to surprising results, but it provides little evidence as to the
true ground states.

We have run numerous Parrinello-Rahman simulations with $2 \le n \le
8$, $1 \le N \le 24$ (and occasionally larger), and various densities,
with the following results (see also Table~\ref{table:energy}).

\begin{table}
\begin{tabular}{ccc|ccc}
$n$ & Energy & State & $n$ & Energy & State\\
\hline
$1$ & $0.04321740\dots$ & $\Z$ & $5$ & $0.17434205\dots$ & $\textup{pc}_2$\\\
$2$ & $0.07979763\dots$ & $A_2$ & $6$ & $0.19437337\dots$ & $\mathcal{P}_6(1.0525\dots)$\\
$3$ & $0.11576766\dots$ & $\textup{pc}_1$ & $7$ & $0.21222702\dots$ & $D_7^+$\\
$4$ & $0.14288224\dots$ & $D_4$ & $8$ & $0.22788144\dots$ & $E_8$
\end{tabular}
\caption{Lowest known energies in dimension $n$ when $\rho=1$.  The
third column specifies the putative ground state, with
``$\textup{pc}_1$'' and ``$\textup{pc}_2$'' standing for phase
coexistence between $D_3$ and $D_3^*$ and between $D_5^+(1.99750\dots)$
and $D_5^+(0.50062\dots)$, respectively.} \label{table:energy}
\end{table}

In two and three dimensions, we observe the previously known ground
states.  In four dimensions, we find the $D_4$ lattice at all
densities, in accordance with the conjecture in Ref.~\cite{TS2}. Thus,
it appears probable that, like $E_8$ and the Leech lattice, the $D_4$
lattice is universally optimal.

In five dimensions, we find different structures.  The $\Lambda_5^2$
lattice was used in Ref.~\cite{CK2} to improve on Bravais lattices at
low density; it consists of parallel translates of $D_4$, repeating
with period $4$.  Parrinello-Rahman simulations rapidly identify and
improve on this structure.  It can be deformed by compressing the
spacing between the parallel copies by some factor.  A local minimum
for energy is achieved for carefully optimized values of the
compression factor, which are between $0.998749\dots$ and $1$ when
$\rho \le 1$ and between $0.25$ and $0.250312\dots$ when $\rho \ge 1$.
Our simulations suggest that these are the true ground states, with the
exception of phase coexistence for $0.99836946\ldots \le \rho \le
1.00163526\ldots$.

These structures fit into the following general family. Let
$$
D_n = \{(x_1,\dots,x_n) \in \Z^n : \textup{$x_1+\dots+x_n$ is even}\}
$$
denote the checkerboard lattice in $\R^n$, and let $D_n^+$ be the union
of $D_n$ with its translation by $(1/2,1/2,\dots,1/2)$.  Let
$D_n^+(\alpha)$ be $D_n^+$ with the last coordinate scaled by a factor
of $\alpha$.  That is,
$$
D_n^+(\alpha) = \{(x_1,\dots,x_{n-1},\alpha x_n) : (x_1,\dots,x_n) \in D_n^+\}.
$$
Then $D_5^+(\alpha)$ is the deformation of $\Lambda_5^2$ with
compression factor $\alpha/2$.  This is not obvious, but it can be
checked by a straightforward computation, and in fact it gives a
substantially simpler construction of $\Lambda_5^2$ than was previously
known [namely, as $D_5^+(2)$].

One noteworthy aspect of these configurations is their anisotropy.  As
the density increases, they experience greater compression along a
distinguished axis than orthogonally to it. However, an even more
surprising phenomenon is that there are formal duality relationships
between these structures. Formal duality is a generalization of the
relationship between a Bravais lattice and its reciprocal lattice (see
Sec.~\ref{section:duality} for more details). Formal duality relates
the energies at densities $\rho$ and $1/\rho$: if $E_\rho$ is the
Gaussian core energy at density $\rho$ for a given structure and
$\widetilde E_\rho$ is that for its formal dual, then
$$
\frac{2E_\rho + 1}{2 {\widetilde E}_{1/\rho} + 1} = \rho.
$$
More generally, formal duality relates the energy of one structure
under a given pair potential to that of the formally dual structure
under the Fourier transform of the potential.

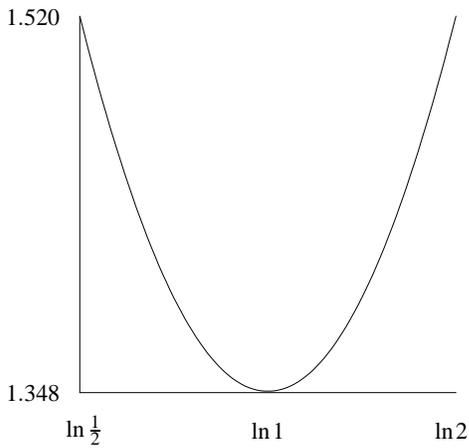
\begin{figure}
\setlength{\unitlength}{1.25cm}
\begin{center}
\begin{picture}(5.5,4.9)(-1,-0.7)
\path(0,0)(4,0)
\path(0,0)(0,4)
\put(-0.45,-0.9){\makebox(1,1){$\ln \frac{1}{2}$}}
\put(1.5,-0.9){\makebox(1,1){$\ln 1$}}
\put(3.45,-0.9){\makebox(1,1){$\ln 2$}}
\put(-1,-0.5){\makebox(1,1){$1.348$}}
\put(-1,3.5){\makebox(1,1){$1.520$}}
\path(0, 4.0067737)(0.040000000, 3.8495461)(0.080000000, 3.6953816)%
(0.12000000, 3.5442873)(0.16000000, 3.3962703)(0.20000000, 3.2513378)%
(0.24000000, 3.1094967)(0.28000000, 2.9707541)(0.32000000, 2.8351168)%
(0.36000000, 2.7025916)(0.40000000, 2.5731855)(0.44000000, 2.4469049)%
(0.48000000, 2.3237566)(0.52000000, 2.2037469)(0.56000000, 2.0868824)%
(0.60000000, 1.9731692)(0.64000000, 1.8626135)(0.68000000, 1.7552214)%
(0.72000000, 1.6509987)(0.76000000, 1.5499513)(0.80000000, 1.4520847)%
(0.84000000, 1.3574044)(0.88000000, 1.2659158)(0.92000000, 1.1776239)%
(0.96000000, 1.0925339)(1.0000000, 1.0106505)(1.0400000, 0.93197849)%
(1.0800000, 0.85652224)(1.1200000, 0.78428608)(1.1600000, 0.71527415)%
(1.2000000, 0.64949037)(1.2400000, 0.58693847)(1.2800000, 0.52762199)%
(1.3200000, 0.47154427)(1.3600000, 0.41870844)(1.4000000, 0.36911743)%
(1.4400000, 0.32277397)(1.4800000, 0.27968054)(1.5200000, 0.23983945)%
(1.5600000, 0.20325278)(1.6000000, 0.16992237)(1.6400000, 0.13984987)%
(1.6800000, 0.11303670)(1.7200000, 0.089484042)(1.7600000, 0.069192878)%
(1.8000000, 0.052163953)(1.8400000, 0.038397789)(1.8800000, 0.027894683)%
(1.9200000, 0.020654709)(1.9600000, 0.016677714)(2.0000000, 0.015963322)%
(2.0400000, 0.016677714)(2.0800000, 0.020654709)(2.1200000, 0.027894683)%
(2.1600000, 0.038397789)(2.2000000, 0.052163953)(2.2400000, 0.069192878)%
(2.2800000, 0.089484042)(2.3200000, 0.11303670)(2.3600000, 0.13984987)%
(2.4000000, 0.16992237)(2.4400000, 0.20325278)(2.4800000, 0.23983945)%
(2.5200000, 0.27968054)(2.5600000, 0.32277397)(2.6000000, 0.36911743)%
(2.6400000, 0.41870844)(2.6800000, 0.47154427)(2.7200000, 0.52762199)%
(2.7600000, 0.58693847)(2.8000000, 0.64949037)(2.8400000, 0.71527415)%
(2.8800000, 0.78428608)(2.9200000, 0.85652224)(2.9600000, 0.93197849)%
(3.0000000, 1.0106505)(3.0400000, 1.0925339)(3.0800000, 1.1776239)%
(3.1200000, 1.2659158)(3.1600000, 1.3574044)(3.2000000, 1.4520847)%
(3.2400000, 1.5499513)(3.2800000, 1.6509987)(3.3200000, 1.7552214)%
(3.3600000, 1.8626135)(3.4000000, 1.9731692)(3.4400000, 2.0868824)%
(3.4800000, 2.2037469)(3.5200000, 2.3237566)(3.5600000, 2.4469049)%
(3.6000000, 2.5731855)(3.6400000, 2.7025916)(3.6800000, 2.8351168)%
(3.7200000, 2.9707541)(3.7600000, 3.1094967)(3.8000000, 3.2513378)%
(3.8400000, 3.3962703)(3.8800000, 3.5442873)(3.9200000, 3.6953816)%
(3.9600000, 3.8495461)(4.0000000, 4.0067737)
\end{picture}
\end{center}
\caption{A plot of $\rho^{-1/2}(2U_\rho + 1)$ as a function of $\ln \rho$, where
$U_\rho$ is the minimal energy attained by the structures $D_5^+(\alpha)$
at density $\rho$.  The reflection symmetry follows from formal duality.}
\label{figure:plot}
\end{figure}

A non-Bravais lattice typically has no formal dual. Thus, it is
remarkable that $D_n^+(\alpha)$ is formally dual to $D_n^+(1/\alpha)$.
(See Sec.~\ref{section:duality} for a proof.)  Formal duality implies
that the high-density ground states in $\R^5$ tend to $D_5^+(1/2)$,
because the low-density ones tend to $D_5^+(2) = \Lambda_5^2$.
Figure~\ref{figure:plot} illustrates the formal dualities within the
$D_5^+(\alpha)$ family of structures.

In six dimensions, the lowest-energy states previously known were the
Bravais lattice $E_6$ for low density and its reciprocal lattice
$E_6^*$ for high density, with a narrow region of phase coexistence in
between.  The lattices remain the lowest-energy states known at extreme
densities, but in between them, our simulations have identified other
candidate ground states.  They are all deformations of the orthogonal
direct sum $D_3 \oplus D_3$ together with its translates by three
vectors, namely,
\begin{align*}
& \left(\frac{1}{2},\frac{1}{2},\frac{1}{2},\frac{1}{2},
\frac{1}{2},\frac{1}{2}\right),\\
& \left(1,1,1,-\frac{1}{2},-\frac{1}{2},-\frac{1}{2}\right),
\textup{ and}\\
& \left(-\frac{1}{2},-\frac{1}{2},-\frac{1}{2},1,1,1\right).
\end{align*}
(These vectors are made up of holes in $D_3$.)  Define
$\mathcal{P}_6(\alpha)$ to be the structure obtained by scaling the
first three coordinates by a factor of $\alpha$ and the last three by
$1/\alpha$, so that volume is preserved.  For certain values of
$\alpha$, these structures improve on $E_6$ and $E_6^*$ for
$0.25384516\ldots \le \rho \le 3.93940925\ldots$, with phase
coexistence for $3.93255017\ldots \le \rho \le 3.94624440\ldots$ and
for the reciprocal range of densities. Unlike $D_5^+(\alpha)$, which
can be viewed as a modification of $D_5^+$ or $\Lambda_5^2$, the
structures $\mathcal{P}_6(\alpha)$ differ more substantially from
previously analyzed structures.  As in five dimensions, however, there
are formal duality relations.  Specifically, $\mathcal{P}_6(\alpha)$ is
formally dual to $\mathcal{P}_6(1/\alpha)$ for each $\alpha$ [of course
$\mathcal{P}_6(1/\alpha)$ is isometric to $\mathcal{P}_6(\alpha)$].

In seven dimensions, we find the $D_7^+(\alpha)$ family of structures
at all densities.  For $0.04660088\ldots \le \rho \le
21.45881937\ldots$, the ground state seems to be $D_7^+$ itself (i.e.,
$\alpha=1$).  For lower densities, we have $\alpha > 1$ and for higher
densities we have $\alpha < 1$.  Unlike the case of $\R^5$, there is no
phase coexistence, because the optimal value of $\alpha$ changes
continuously as a function of density.  The low-density limit of
$D_7^+(\alpha)$ is $D_7^+(\sqrt{2})$, which is the same as the
$\Lambda_7^3$ structure studied in Ref.~\cite{CK2}. By formal duality,
the high-density limit is $D_7^+(1/\sqrt{2})$.

Finally, in eight dimensions our simulations provide further evidence
that $E_8$ (i.e., $D_8^+$) is universally optimal.  Parrinello-Rahman
simulations are by no means limited to eight dimensions, and in fact,
we have numerical results in as many as twelve dimensions.  These
results, together with a more extensive analysis of the structures
presented here, will appear elsewhere \cite{CKS}.

Within the $D_n^+(\alpha)$ family of structures for $1 \le n \le 8$,
the best energy at low density (i.e., the best sphere packing) is
obtained when $\alpha = \sqrt{9-n}$.  For $n \le 4$, the
$D_n^+(\sqrt{9-n})$ configuration is inferior to previously known
sphere packings. However, for $5 \le n \le 8$ it achieves the highest
sphere packing density currently known.  Note that $D_6^+(\sqrt{3})$ is
the $\Lambda_6^2$ packing, which has the same packing density as $E_6$
but is slightly inferior in the Gaussian core model at low densities.
We have no conceptual explanation for why the six-dimensional behavior
is subtly different from that in five, seven, or eight dimensions.

It is also interesting to examine the $\alpha = 1$ case.  It seems that
the $D_5^+$ and $D_6^+$ structures are not local optima for the
Gaussian core model at any density. By contrast, $D_7^+$ appears to be
the ground state over a large range of densities, and $D_8^+$ is almost
certainly the ground state at all densities.  It is possible that
$D_9^+$ is also universally optimal: so far we have not explored this
case as thoroughly as those in lower dimensions, but we have not yet
found any structure that beats $D_9^+$ at any density.  We will examine
this issue elsewhere \cite{CKS}.  Universal optimality cannot hold for
$D_n^+$ with $n \ge 10$, because these packings are not even optimal
sphere packings.

\section{Formal duality} \label{section:duality}

Recall that Poisson summation relates the sum of a function over a
Bravais lattice to the sum of its Fourier transform over the reciprocal
lattice.  Specifically, given a sufficiently well-behaved function $f
\colon \R^n \to \R$ (for example, a Schwartz function) and a Bravais
lattice $\Lambda \subset \R^n$,
$$
\sum_{x \in \Lambda} f(x) = \frac{1}{\vol(\R^n/\Lambda)}
\sum_{y \in \Lambda^*} \widehat{f}(y).
$$
Here, $\vol(\R^n/\Lambda)$ denotes the volume of a fundamental domain
of $\Lambda$ and we normalize the Fourier transform and reciprocal
lattice by
$$
\widehat{f}(y) = \int_{\R^n} f(x)\, e^{2\pi i \langle x,y \rangle} dx
$$
and
$$
\Lambda^* = \{y \in \R^n: \langle x,y\rangle \in \Z
\textup{ for all $x \in \Lambda$}\}.
$$

Formal duality is a generalization of Poisson summation to certain
special structures that are not Bravais lattices. To state it
correctly, it is important to view the sum
$$
\sum_{x \in \Lambda} f(x)
$$
not simply as a sum over points in $\Lambda$, but rather as a sum over
vectors between points in $\Lambda$.  For a Bravais lattice these
notions are exactly the same, but in more general cases they are not.
(In fact, Poisson summation cannot be generalized from the first point
of view \cite{C}.) For example, consider a periodic configuration given
by the union of $N$ disjoint lattice translates
$\Lambda+v_1,\dots,\Lambda+v_N$. Then the analog of summing over the
lattice is
$$
\frac{1}{N} \sum_{j=1}^N \sum_{k=1}^N \sum_{x \in \Lambda} f(x+v_j-v_k).
$$
Call this sum the \emph{average pair sum} of $f$ over the
configuration. It is the average over all points in the configuration
of the sum of $f$ over all vectors from it to other points.  (Thus, it
is independent of how the configuration is decomposed into translates
of Bravais lattices.)

Suppose $\mathcal{P}$ and $\mathcal{Q}$ are particle arrangements,
where $\mathcal{P}$ has particle density $\delta$ and $\mathcal{Q}$ has
particle density $1/\delta$. We say $\mathcal{P}$ and $\mathcal{Q}$ are
\emph{formal duals} if for every Schwartz function $f$, the average
pair sum of $f$ over $\mathcal{P}$ is $\delta$ times that for
$\widehat{f}$ over $\mathcal{Q}$.  Poisson summation shows that this
definition generalizes the case of reciprocal Bravais lattices.  If
$\mathcal{P}$ is formally dual to itself, we call it \emph{formally
self-dual}, and if it is formally dual to an isometric copy of itself,
we call it \emph{formally isodual}.  For example, the triangular
lattice in the plane is not self-dual, since its reciprocal lattice is
a rotated copy of itself, but it is isodual.

For any periodic configuration, one can always write the average pair
sum of $f$ in terms of $\widehat{f}$, by using a generalized Poisson
summation formula: for a Bravais lattice $\Lambda$ and translation
vector $v$,
$$
\sum_{x \in \Lambda} f(x+v) = \frac{1}{\vol(\R^n/\Lambda)}
\sum_{y \in \Lambda^*} e^{-2\pi i \langle v,y \rangle} \widehat{f}(y).
$$
(In fact, the right side is the Fourier expansion of the left side as a
function of $v$ that is periodic modulo $\Lambda$.)  The average pair
sum of $f$ over $\Lambda+v_1,\dots,\Lambda+v_N$ then becomes
$$
\frac{N}{\vol(\R^n/\Lambda)} \sum_{y \in \Lambda^*} \widehat{f}(y)
\left|\frac{1}{N}{\sum_{j=1}^N
e^{2\pi i \langle v_j,y\rangle}}\right|^2.
$$
The factor of $N/\vol(\R^n/\Lambda)$ is the density of the
configuration, so the question becomes whether the remaining sum is the
average pair sum of $\widehat{f}$ over some periodic structure. Except
when $N=1$, it usually is not: for example, the coefficient of
$\widehat{f}(y)$ is generally irrational.  Formal duality only arises
in exceptional cases.  It is not obvious when a configuration has a
formal dual or, if it does have one, what the formal dual is.

\begin{proposition} \label{proposition:dnplus}
The $D_n^+$ structure is formally self-dual when $n$ is odd or a
multiple of four. When $n$ is even but not a multiple of four, $D_n^+$
is formally isodual.
\end{proposition}

When $n$ is even, $D_n^+$ is a Bravais lattice, whose reciprocal
lattice is $D_n^+$ if $n$ is a multiple of four and $D_n^+(-1)$
otherwise. When $n$ is odd, $D_n^+$ is not a Bravais lattice and the
duality is more subtle.

\begin{proof}
Let $v = (1/2,1/2,\dots,1/2)$, so $D_n^+$ is the union of $D_n$ and
$D_n+v$. Then the average pair sum of $f$ over $D_n^+$ is
$$
\frac{1}{4} \sum_{y \in D_n^*} \widehat{f}(y)
\left|1 + e^{2\pi i \langle v,y \rangle}\right|^2 =
\sum_{y \in D_n^*} \widehat{f}(y) \frac{1+\cos (2\pi \langle v,y \rangle)}{2}.
$$
The reciprocal lattice $D_n^*$ consists of four translates of $D_n$, by
the vectors $0$, $v$, $(0,0,\dots,0,1)$, and
$(1/2,1/2,\dots,1/2,-1/2)$.  In each of these four cases, the inner
product $\langle v,y \rangle$ is easily understood.  In the first case,
it is an integer, in the second it is an integer plus $n/4$, in the
third it is an integer plus $1/2$, and in the fourth it is an integer
plus $(n-2)/4$.

When $n$ is odd, this yields weights of $1$, $1/2$, $0$, and $1/2$
multiplying $\widehat{f}(y)$ in the four cases.  Because $n$ is odd,
the vector $(1,1,\dots,1,0)$ is in $D_n$ and hence translating $D_n$ by
$(1/2,1/2,\dots,1/2,-1/2)$ is equivalent to translating it by $-v$.
Thus, the average pair sum for $f$ is simply
$$
\frac{1}{2} \sum_{x \in D_n} \left(2\widehat{f}(x) +
\widehat{f}(x+v) + \widehat{f}(x-v)\right),
$$
which is the same as the average pair sum for $\widehat{f}$ over
$D_n^+$. It follows that $D_n^+$ is formally self-dual when $n$ is odd.

An analogous computation works in the even case, or it can be verified
more simply using the fact that $D_n^+$ is then a Bravais lattice.
\end{proof}

\begin{lemma} \label{lemma:linear}
If $\mathcal{P}$ and $\mathcal{Q}$ are formally dual structures in
$\R^n$, and $T \colon \R^n \to \R^n$ is an invertible linear
transformation, then $T \mathcal{P}$ and $(T^t)^{-1} \mathcal{Q}$ are
formally dual.
\end{lemma}

Here $T^t$ denotes the adjoint operator with respect to the inner
product (i.e., the transposed matrix).

\begin{proof}
To compute the average pair sum for $f$ on $T \mathcal{P}$, simply
compute it for the composition $f \circ T$ on $\mathcal{P}$.  The
Fourier transform of $f \circ T$ is $(\det T)^{-1} \big(\widehat{f}
\circ (T^t)^{-1}\big)$.  Now applying formal duality for $\mathcal{P}$
and $\mathcal{Q}$ completes the proof.
\end{proof}

It follows immediately that $D_n^+(\alpha)$ is formally dual to
$D_n^+(1/\alpha)$ when $n$ is odd or a multiple of four: apply
Lemma~\ref{lemma:linear} with $\mathcal{P}=\mathcal{Q}=D_n^+$ and with
$T$ being the map that multiplies the last coordinate by $\alpha$.

The formal isoduality of $\mathcal{P}_6(\alpha)$ is proved similarly.
For $\alpha=1$, formal self-duality follows from a calculation much
like the proof for Proposition~\ref{proposition:dnplus}.  Then
Lemma~\ref{lemma:linear} implies that $\mathcal{P}_6(\alpha)$ is
formally dual to $\mathcal{P}_6(1/\alpha)$, but of course the two
configurations are isometric.

In the literature, formal duality is usually understood to refer only
to radial functions $f$ (see, for example, p.~185 of
Ref.~\cite{SPLAG}).  That is a weaker condition, which depends only on
the radial pair correlation functions of the structures.  We have
defined a stronger version of formal duality in this paper, without
that restriction, because the stronger version in fact holds for the
structures we find in our simulations.  (Furthermore, it behaves
better.  For example, the proof of Lemma~\ref{lemma:linear} breaks down
in the radial case, because $f$ and $f \circ T$ will generally not both
be radial.) However, for the discussion in Sec.~\ref{section:results},
the pair potential is isotropic, so only the radial version of formal
duality is needed. Note also that radial symmetry erases the
distinction between formal self-duality and formal isoduality.

\section{Conclusions and discussion}

We have used Parrinello-Rahman dynamics to identify ground states in
dimensions that were previously beyond the reach of simulation. This
approach is effective because it adapts to whichever underlying Bravais
lattice is most favorable.  That means it probably offers little
advantage in detecting disordered states or even phase coexistence, but
it is appropriate whenever one anticipates a high degree of symmetry.

The formal duality relations are the most noteworthy consequence of our
simulations.  Such relations occur only rarely for structures other
than Bravais lattices, and it is far from obvious why they arise here.
A remarkable possibility is that all periodic ground states of the
Gaussian core model, in any dimension, may occur in formally dual
pairs.  If true, this hypothesis deserves a more conceptual explanation
than a case-by-case calculation, and it suggests that the model
possesses deeper symmetries than are currently understood.  Even if it
is false, there must be a reason why formal duality arises so
frequently in low dimensions.

The families of structures studied in this article have much broader
applicability than just to the Gaussian core model.  We believe that
they will minimize many other repulsive potential functions, such as
inverse power laws, although we have done relatively little
experimentation in this direction.

One reason for our focus on the Gaussian core model is that it is the
natural setting for studying universal optimality \cite{CK1}.  The
known universal optima include some of the most symmetrical and
beautiful geometrical configurations, with connections to many other
topics such as sporadic finite simple groups and exceptional Lie
algebras.  Relatively few universal optima are known and any new
examples are of interest. In this article, we have described simulation
evidence that $D_4$ is likely universally optimal and that $D_9^+$ may
be.  Both cases are surprising: the analog in spherical geometry of the
universal optimality of $D_4$ turned out to be false \cite{CCEK}, and
there have not even been any previous hints that $D_9^+$ might be
universally optimal.

Our results also offer insight into the complexity of ground states.
One measure of the complexity of a lattice is the number of Bravais
lattice translates required to generate it (in crystallographic terms,
the minimal size of a particle basis).  Bravais lattices have
complexity $1$, while disordered structures can be considered to have
infinite complexity.  The hexagonal close-packing has complexity $2$,
while the structures introduced here have complexity $2$ (in five and
seven dimensions) and $4$ (in six dimensions).

Do the complexities of ground states grow with dimension?  The
Torquato-Stillinger decorrelation conjecture \cite{TS1} suggests that
they do grow and eventually become infinite.  If so, how quickly do
they grow?  There is a striking example in ten dimensions (the Best
packing \cite{SPLAG}, which is the densest sphere packing known in
$\R^{10}$, and which has complexity $40$), but other low-dimensional
ground states for repulsive potentials that have been reported in the
literature typically have much smaller complexity, with the exception
of phase coexistence.

In up to eight dimensions, our results for the Gaussian core model
suggest that the ground states may indeed have low complexity for most
densities. The structures we have identified seem difficult to improve,
even if we allow the algorithm the freedom of substantially higher
complexity, and we suspect that they are the true ground states. Of
course, we cannot rule out the possibility that extraordinarily
high-complexity states offer tiny improvements, but we consider it
unlikely.

We conclude with a computational challenge regarding simulation in high
dimensions.   It is undoubtedly impossible to carry out effective
simulations in extremely high dimensions, but where is the threshold
for feasibility?  For example, is it possible in $24$ dimensions?  Many
remarkable phenomena in mathematics and information theory (such as the
Leech lattice \cite{SPLAG}) occur there, and reliable simulation
results would be very interesting.

\section*{Acknowledgments}

We thank Salvatore Torquato and Burkhard D\"unweg for helpful
discussions, an anonymous referee for useful suggestions, and Frank
Vallentin for supplying his implementation \texttt{shvec} of the
Fincke-Pohst algorithm. A.K.\ was supported in part by the National
Science Foundation under Grant No.\ DMS-0757765 and by a grant from the
Solomon Buchsbaum Research Fund. A.S.\ thanks Microsoft Research New
England for its hospitality and was supported in part by the Deutsche
Forschungsgemeinschaft under Grant No.\ SCHU 1503/4-2.

\end{document}